\documentclass{article}
\usepackage{setspace}
\usepackage{graphicx}
\usepackage[super]{natbib}
\usepackage{amssymb, latexsym}
\usepackage{amsmath} 
\usepackage{amsthm}
\usepackage{bm}
\usepackage[mathscr]{eucal}
\usepackage{enumerate}
\usepackage{multirow} 
\usepackage[center]{caption}
\usepackage{musicography}
\usepackage{float}

\usepackage{geometry}

\usepackage{authblk}
\usepackage{abstract}

\begin{document}
	
	\title{Modeling Temporal Lobe Epilepsy during Music Large-Scale Form Perception using the Impulse Pattern Formulation (IPF) Brain Model}
	
	\author{Rolf Bader}
	\affil{ Institute of Musicology\\ University of
		Hamburg\\ Neue Rabenstr. 13, 20354 Hamburg, Germany\\
	}
	\date{\today}
	
	\twocolumn
	[
	\begin{@twocolumnfalse}
		
		\maketitle
		\begin{abstract}
		Musical large-scale form is investigated using an Electronic Dance Music (EDM) piece fed into a Finite-Difference Time Domain (FDTD) physical model of the cochlear which again inputs into an Impulse-Pattern Formulation (IPF) brain model. In previous studies, experimental EEG data showed an enhanced correlation between brain synchronization and the musical piece's amplitude and fractal correlation dimension in good agreement with a FitzHugh-Nagumo oscillator model\cite{Sawicki2022}. Still, this model cannot display temporal developments of large-scale forms. The IPF Brain model also shows a high correlation between cochlear input and brain synchronization at the gamma band range around 50 Hz, but also a strong negative correlation for low frequencies, associated with musical rhythm, during time frames of low cochlear input amplitude. Such high synchronization corresponds to temporal lobe epilepsy, often associated with creativity or spirituality. Therefore, the IPF Brain model suggests those conscious states to happen in times of low external input at low frequencies where isochronous musical rhythms are present.
		\end{abstract}

\end{@twocolumnfalse}
]		
		
		\vspace{1cm}
		
		
		\renewcommand \thesection{\Roman{section}}
		\renewcommand \thesubsection{\Alph{subsection}} 
		\renewcommand \thesubsubsection{\arabic{subsubsection}} 
		
	\section{Introduction}
		
	Musical large-scale forms, the temporal organization of a musical piece over its whole length, e.g., the song structure of verse and refrain, Electronic Dance Music (EDM) tension build-up and relaxation, or sonata song structures are poorly researched in terms of brain dynamics\cite{Hartmann2014}\cite{Hartmann2020}. It is widely accepted that music is organized according to Gestalt principles\cite{Bregmann1990}\cite{Leman1997}, often with hierarchical structures, psychologically\cite{Deliege2014}, or in terms of music theory\cite{Lerdahl1990}, and subject to long-term memory in music\cite{Cuddy2018}.

	The processing of sound in the brain has historically been taken as a bottom-up process, starting from the transition of sound in the cochlear into neural spikes\cite{Bader2015} further processed in the auditory pathway\cite{Pressnitzer2004}, showing coincidence detection\cite{Bader2018b}, tonotopy, interaural level and time difference detection, or pitch perception\cite{Cariani2001}\cite{Bader2017}\cite{Bader2021}, next to others. Still, even within the auditory cortex, multiple bottom-up and top-down connections are present\cite{Hawkins1996} and so viewing the auditory pathway as a complex, self-organizing neural network seems more appropriate.

	Such a self-organizing view is standard in terms of cortical processing and existing neural models vary in terms of complexity and scaling\cite{Kacprzyk2015}. Only a few brain models try to understand the brain with simple principles like the free-energy principle\cite{Friston2013} assuming adaptation of the brain to external surprises, a global workspace view\cite{Baars2013} assuming synchronization and de-synchronization of brain parts over time, or a Synergetic approach of Gestalt perception\cite{Haken2008}. 

	Machine-learning models also view conscious content as a process of complex, often nonlinear interactions between single neurons leading to a heuristic and coherent Gestalt, like connectionist\cite{Grossberg1976a}, \cite{Grossberg1976b} music models\cite{Gjerdingen1990}\cite{Briot2020} also when analyzing large musical databases as Computational Phonogram Archiving\cite{Bader2019}\cite{Blass2019} for streaming platforms or archives in general\cite{Blass2020}.

	The self-organizing idea is also reflected in the idea of 50-millisecond intervals of organized neural spatiotemporal patterns followed by short, chaotic disturbances associated with olfactory\cite{Kozma2016} or auditory\cite{Ohl2001}\cite{Ohl2016} conscious content. Enlarging the picture to interactions between subjects results in the idea of a self-organizing society of brains\cite{Freeman2014} or including cultural artifacts and nature in a Physical Culture Theory\cite{Bader2018a}.

	Incorporating brains with cultural objects asks for a general framework which has been proposed as Impulse Pattern Formulation (IPF) at first developed for musical instruments. As physically the only force we experience - next to gravity - is electricity it is straightforward to formulate acoustic as well as electric nervous spike impulses as being of the same nature. The IPF then takes a viewpoint neuron or musical instrument part from which an impulse is sent out to several other neurons, musical instrument parts, or any kind of object. This impulse is processed, damped, and returns back to the viewpoint object\cite{Bader2013}\cite{Linke2019a}\cite{Bader2021}. Such an iterative, nonlinear-dynamical process is scale-free, capable of modeling sudden phase changes, and includes convergent, bifurcating, complex, and chaotic states. Although often modeling a system with only very few nodal points, the IPF has already been shown to be of high precision in musical instrument applications\cite{Linke2022}\cite{Linke2021b}\cite{Linke2019b} as well as in rhythm perception and production\cite{Linke2021a}. 

	The model was also formulated as a brain model\cite{Bader2022} with neural adaptation and plasticity non-trivially finding the concentration of inhibitory vs. excitatory neurons of 10-20\% as a maximum of possible system convergence, a maximum reflection strength around the usual time for Event-Related Potentials (ERP), as well as a decay of memory in the system corresponding to short-time memory. These general findings are strongly pointing to the validity of the model.
	
	Investigating brain dynamics of large-scale musical form has already shown musical tension to be represented in increasing synchronization of brain parts\cite{Hartmann2014}\cite{Hartmann2020}\cite{Sawicki2022}. Such findings correspond to synchronization caused by expectancy\cite{Buhusi2005} towards a climax. A main finding is that brain synchronization is strongest around 50 Hz, so in the gamma band of brain dynamics.

	An interesting aspect of the IPF Brain model is the presence of convergence, i.e., the complete synchronization of the neurons in the model. This corresponds to epilepsy\cite{Gerster2020} which is not the usual and a very dangerous brain state. Still, partial synchronization in the brain is well known as chimera states\cite{Omelchenko2013}. Also in terms of the auditory cortex, temporal lobe epilepsy has often been reported\cite{McCrae2012} and associated with spiritual or meditative states of mind. Such states are naturally associated with larger time spans than pitch or rhythm and are subject to large-scale forms.

	The present paper applies the IPF Brain model as suggested before\cite{Bader2022} to the case of an Electronic Dance Music (EDM) piece \emph{One Mic} of the Rapper NAS which has been investigated before in EEG experiments\cite{Hartmann2014}\cite{Hartmann2020} and by using a FitzHugh-Nagumo dynamical brain model\cite{Sawicki2022}. Strong correlations between the large-scale form of the model, associated with the sound amplitude and fractal correlation dimension as a measure of musical density\cite{Bader2013} have been found experimentally and in the model with in both cases a maximum of synchronization in the gamma band of brain dynamics around 50 Hz. Still, due to the FitzHugh-Nagumo model not dissolving the temporal development of the brain dynamics, a deeper understanding of the reason for this frequency-dependency of synchronization asks for a model capable of such a temporal resolution, which is performed using the IPF Brain model in this paper.

	The paper first introduces the method used. As in the previous paper, the input to the Brain IPF is the output of a Finite-Difference Time-Domain (FDTM) physical model of the cochlear to which the musical sound is used as an input. The cochlear input to the Brain IPF results in the development of both the IPF system parameter g, the state of the viewpoint neuron, as well as the adaptation of the reflection points, the neurons. To estimate the influence of the amount of neurons used in the model, N=50, 100, and 200 neurons are taken, showing the independence of the model with respect to this parameter. In the method section, the post-processing of these modeled parameters is discussed, especially the Kuramoto order parameter to measure the synchronization of neurons and the correlation of this parameter with the cochlear input. The result section then mainly concentrates on this correlation and discusses the reason for the model behavior. In the conclusions, an overview of different conscious states with respect to the synchronization frequency concludes the paper.

		\section{Method}
		
		\subsection{Cochlear Model}
		
		The present model assumes a differential equation of a membrane
		
		\begin{equation}
			\frac{K(x)}{\mu(x)}\frac{\partial^2 u}{\partial x^2} - d(x) \frac{\partial u}{\partial t} = \frac{\partial^2 u}{\partial t^2} + f(t)\ ,
		\end{equation}
		
		with BM displacement u along a one-dimensional axis x, BM stiffness $K(x) = 2 \times 10^9 e^{-3.4 x} dyn / cm^3$ changing along x and linear density $\mu(x) = m / A(x)$ with mass over area again changing along the BM and $A(x) = .1 cm \times (.1 cm + 0.02 cm \times x / l)$ with BM length $l = 3.5 cm$ taking the slight widening of the BM over its length into account.
		
		A 1-D model is sufficient to model a BM as shown by \cite{Babbs2011} comparing a 1D and 2D model based on the anisotropy of the BM as discussed by \cite{Liu2008}. Here it is found numerically and based on experimental data that the inclusion of a second dimension contributes less than 1\% of the results already obtained by a 1-D model. This is reasonable as the BM has dimensions of about 3.5 cm in length but only about 1 mm in width and is therefore more a rod than a membrane. Also the Young's modulus in y-direction is only about 10\% of that in x-direction \cite{Babbs2011} and does therefore not add considerably to the overall BM movement. 
		
		To confirm this finding, a 2-dimensional model was built
		
		\begin{equation}
			\frac{K_x(x)}{\mu(x,y)}\frac{\partial^2 u}{\partial x^2} + \frac{K_y(x)}{\mu(x,y)}\frac{\partial^2 u}{\partial y^2} - d(x,y) \frac{\partial u}{\partial t} = \frac{\partial^2 u}{\partial t^2} + f(t)\ ,
		\end{equation}
		
		with Young's modulus in x-direction $K_x(x) =  2 \times 10^9 e^{-3.4 x} dyn / cm^3$ as above and in y-direction $K_y(x) = .1 \times K_x(x)$ according to the literature \cite{Babbs2011}. The linear density $\mu(x,y) = mu(x)$ of the 1D model, which holds also for damping $d(x,y) = d(x)$. The model consisted again of 48 modal points in x-direction and 6 nodal points in y-direction. The boundary conditions were again simply supported. Still the results did not differ considerably as shown below comparing the tuning curves of the 1D model in Fig. 3 and those of the 2D model shown in Fig. 4. For sake of simplicity, also taking the computational cost into consideration the 1D model was then used further.
		
		The Electronic Dance Music (EDM) piece \emph{One Mic} by the Rapper NAS was used as a sound input to the cochlear model. The FDTM sample rate was 192 kHz to ensure model stability. The output is a set of spikes time points $S^i_B$ at each of the 24 Bark bands B with i = 1,2,3...$N_B$, where $N_B$ is the maximum number of spikes at Bark band B. Each $S^i_B$ has an associated amplitude $A^i_B$. Although single spikes have a more or less uniform amplitude, the cochlear nerve fiber output of the model sums many spikes at each position $S^i_B$. Therefore $A^i_B$ represents the amount of spikes or the output strength.
		
		As input to the IPF Brain model all output as accumulates with respect to 20 ms time intervals, corresponding to $f_max =$ 500 Hz which then is the time constant of the IPF Brain model discussed below. Therefore, the frequency range of interest in the brain up to about 200 Hz is well represented. The resulting cochlear output time series then is denoted
		
		\begin{equation}
			C_t = \sum_{B=1}^{24} \sum_{i=1}^{f_{max}\times T_{max}} G(A^i_B) H(S^i_B) \ ,
		\end{equation}
		
		where $T_{max}$ = 266 seconds, the length of the musical piece, and the function G and H detect if the respective spike is within the respective time window and gives $G(A^i_B) = A^i_B$ and $H(S^i_B) = 1$ if so. For further correlation with the IPF Brain model output, $C_T$ is calculated as the time series averaged over time windows of 1 second. The unitless mean amplitude of the $\overline{C_t}$ = 0.000118.
		
		\subsection{Brain Impulse Pattern formulation (IPF)}
		
		The brain is modeled using N = 50 neurons. Each neuron is a reflection point, returning impulses from a viewpoint neuron. The system state of the viewpoint neuron is g, which represents a time period and amplitude strength. Each reflection neuron i has a damping $\alpha_i$. The IPF then is
		
		\begin{equation}
			g_t = \frac{1}{N} \left|\sum_{i=1}^{N} P_i \ln \alpha_t^i\ g_{t-i}\right| + w_2\ g^I_t\ .
			\label{IPF_Brain}
		\end{equation}
		
		Here the viewpoint neuron is i=1 the reflections come from neurons i=2,3,4,...N. P$_i$ are the polarizations of the neuron, where P$_i$ = 1 is an excitatory and P$_i$ = -1 is an inhibitory neuron. Throughout the paper a relation of 10\% of inhibitory neurons is used. Note that the sum of all reflections is normalized using the amount of neurons N. The model is discrete with time steps t=0,1,2,3... Therefore, the earlier states of the viewpoint neuron, which this neuron has sent out to the other neurons, are returning after a delay in a damped and polarized form.
		
		For a deeper discussion of the model see\cite{Bader2022}.
		
		\subsection{Plasticity Model}
		
		The plasticity of each neuron is calculated for each time step t. Plasticity means a change in the damping parameter $\alpha_i$, where each time step t then might have a different damping $\alpha_t^i$. Note that in the IPF reasoning, the damping is originally $1/\alpha$, which, for the sake of convenience, is skipped, now using $\alpha$ instead.
		
		For each time step, the new damping is calculated like
		
		\begin{equation}
			\alpha_t^i = \left|\alpha_{t-1}^i + w_1 \ln(1+(g_{t-i} - g_t))\right| \ .
			\label{IPF_plasticity}
		\end{equation}
		
		If the reflection point neuron t-i has the same value $g_{t-i}$  as the viewpoint neuron value $g_t$ the logarithm becomes zero, and no change in the damping $\alpha_{t^i}$ happens. If the reflection point neuron t-i has a larger value than the viewpoint neuron, the logarithm becomes larger than zero, and $\alpha_i$ increases. Otherwise, the logarithm assures a negative influence, and $\alpha_i$ decreases. The plasticity process is generally modeled using a constant $w_1$. Therefore, plasticity can be switched off in the model by using $w_1 = 0$. To examine different model behavior, $w_1$ will systematically be altered, shown below. Again, the absolute value of $\alpha$ is used, not allowing negative or complex values. This, again, does not change the model behavior due to the logarithms used. Still, positive values are more convenient. Indeed, negative arguments of the logarithm in the simulations shown below appear very rarely and are additionally suppressed by using the absolute value.
		
		\subsection{External Musical Instrument Input IPF}
		
		Like with a previous study\cite{Sawicki2022} the Electronic Dance Music (EDM) piece \emph{One Mic} of the Rapper NAS has been used. 
		
		\subsection{Detection of System Behavior}
		
		Each neuron has its own time series
		
		\begin{equation}
			g^i_t = P_i \ln \alpha_t^i\ g_{t-i}
		\end{equation}
		
		when reflecting back to the central neuron, as can be seen in Eq. \ref{IPF_Brain}, where $g^i$ is the i$^{th}$ neuron at time point t with a certain $\alpha^i_t$ at that time point. These time series are Fourier analyzed with time windows of one second, resulting in $F^i_T(f)$, where $0 \leq T \leq 266$ seconds, the length of the musical piece. All IPF simulations in this paper are performed with a maximum of $f_{max}$ = 500 Hz, therefore 1 $\leq$ f $\leq$ 500 Hz. The time series of the central neuron will be labeled below as simply $g$.
		
		Synchronization is measured, as in a previous paper, using the Karumoto order parameter
		
		\begin{equation}
			K_T(f) = \frac{1}{N} \left| \sum_{i=1}^{N} e^{\imath\ \theta^i_T(f) } \right| \ ,
		\end{equation}
		
		where $\theta^i_T(f)$ is the phase of the i$^{th}$ neuron at time interval T of frequency f taken from $F^i_T(f)$. N is the amount of neural reflection points, in this study N=50, N=100, and N=200.
		
		The synchronization order parameter $K_T(f)$ is time-dependent. To estimate the overall synchronization strength, from $K_T(f)$ a time-averaged mean $K(f)$ is calculated. 
		
		Also, the correlation of $K_T(f)$ with the cochlear input time series $C_T$ is performed, leading to $K^C(f)$, an estimation of the frequency-dependency of the correlation strength of the synchronization with the time series.

		\section{Results}
		
		To estimate the influence of the amount of neurons used, analysis was performed for N=50, like in a previous study, as well as N=100, and N=200. 
	
	\subsection{System behavior}
	
\begin{figure}
			\centering
			\includegraphics[width=1\linewidth]{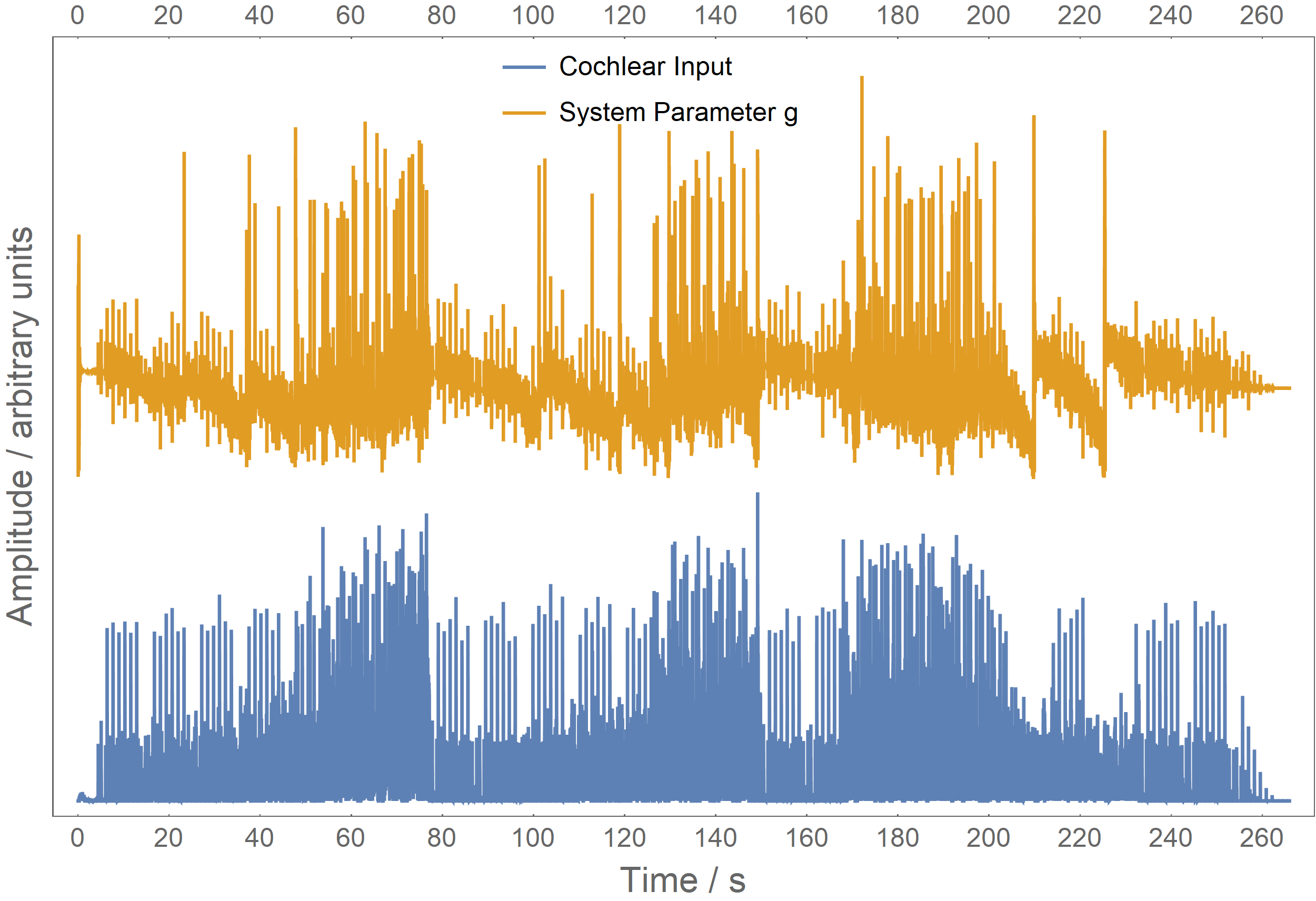}
			\caption{Time series of cochlear input $C_t$ (blue) and system parameter g (yellow) over the 266 seconds of the musical piece \emph{One Mic} by NAS, using N=100 reflection points. Both time series correlate with 0.38.}
			\label{fig:cochlearinputg}
		\end{figure}

	At first, the system parameter g is expected to follow the cochlear input systematically. Fig. \ref{fig:cochlearinputg} shows the cochlear input $C_T$ at the bottom (blue) and the system parameter g of the central neuron (yellow) at the top. The simulation was performed for N=50, N=100, and N=200 neural reflection points, where the figure shows the N=100 case. Visually, the time series correspond well. The correlations are 0.37 for N=50, 0.38 for N=100, and 0.40 for N=200. This is the first example of a high independence of the results with respect to the amount N of neural reflection points.

		\subsection{Gamma band synchronization strength}
		
		The IPF Brain model is expected to reproduce experimental data. It was found that the strongest correlation between brain synchronization and musical sound input is in the gamma band around 50 - 80 Hz. Below and above the gamma band, synchronization is still there, but decreases considerably.

		\begin{figure}
			\centering
			\includegraphics[width=1\linewidth]{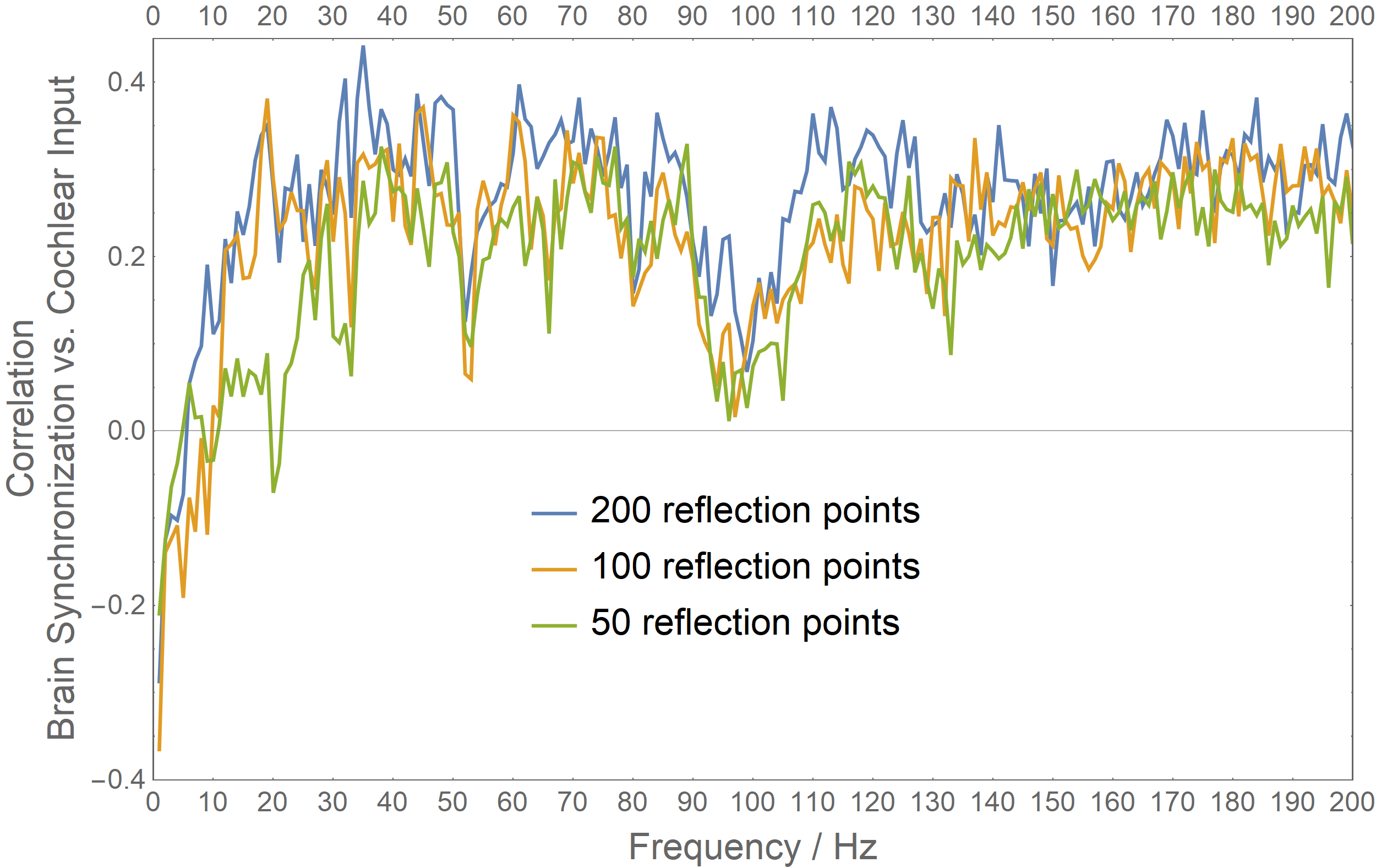}
			\caption{Frequency-dependent correlation of cochlear input with Kuramoto synchronization $K^C(f)$. As expected, the gamma band around 50-80 Hz shows strongest correlation, while correlation decreases to lower and higher frequencies. Very low frequencies have negative correlations, pointing to a conversion of the system during low sound input time windows.}
			\label{fig:synchronizationvsfrequency}
		\end{figure}
		
		Fig. \ref{fig:synchronizationvsfrequency} shows the $K^C(f)$, the correlation of the Kuramoto order parameter with cochlear input $C_T$ for frequencies up to 200 Hz for the three reflection point N=50, N=100, and N=200. As expected, there is a peak of $K^C(f)$ around 50 Hz with decreasing correlation both above 50 Hz, up to about 90-100 Hz, and below 50 Hz. This is consistent through all N=50, 100, and 200, again pointing to the independence of the choice of N. Still, the correlation for N=50 is slightly lower than for N=100 or N=200, a tendency already found above.

\begin{figure}
		\centering
		\includegraphics[width=1\linewidth]{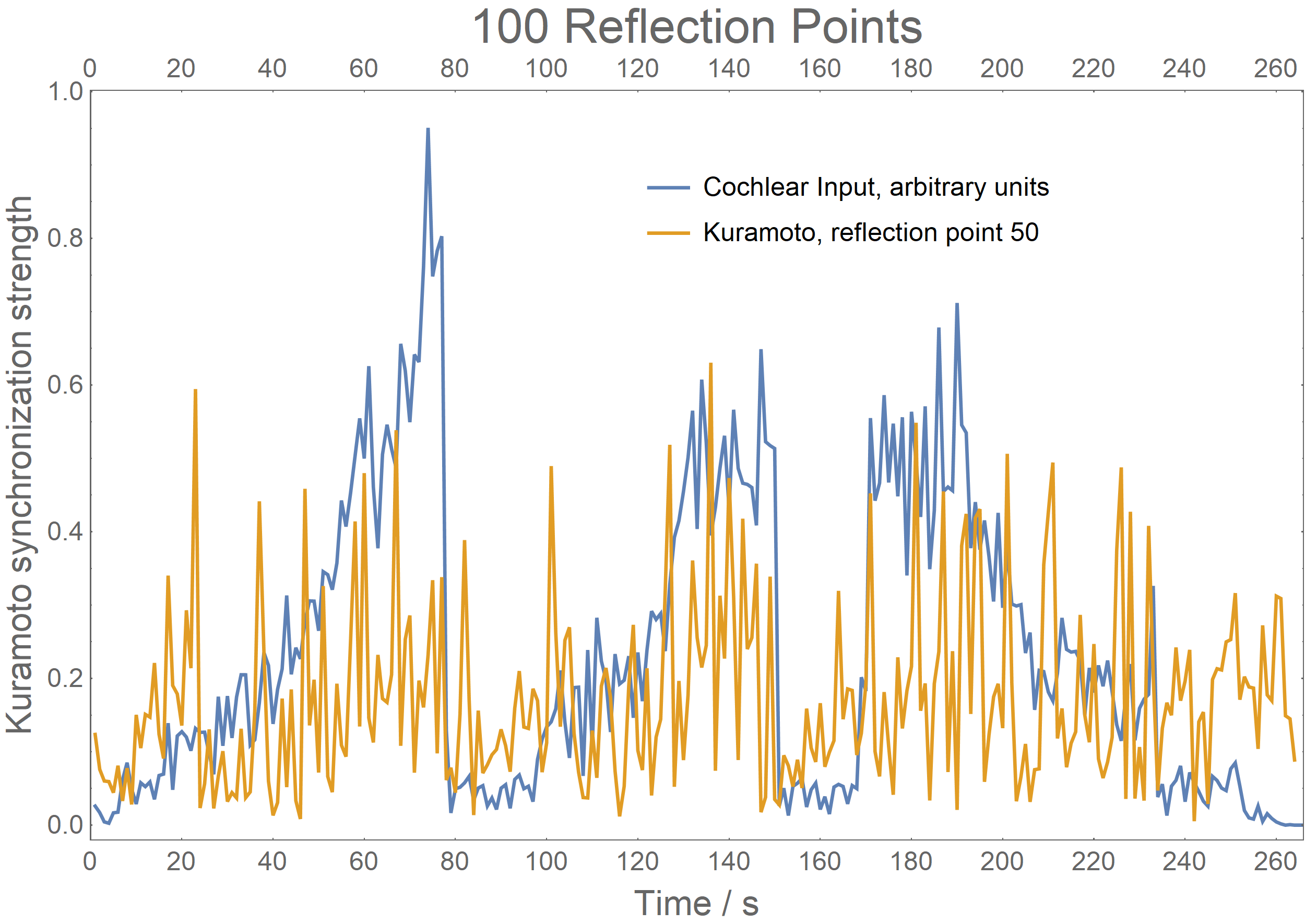}
		\caption{Kuramoto order parameter $K_T(f)$ of 100 neural reflection points at a frequency of 50 Hz (yellow) and cochlear input $C_T$ (blue). Mainly, during windows of strong cochlear input corresponds to large Kuramoto synchronization and vice versa.}
		\label{fig:synchronizationvstime10050}
		\end{figure}

	As an example of the correspondence between $K^C(f)$ and $C_T$ in Fig. \ref{fig:synchronizationvstime10050} the time series of $C_T$ and $K^C(50)$, so for 50 Hz for the N=100 case is shown. Clearly, synchronization follows the amplitude of the cochlear input, the musical piece, nicely. The synchronization is not as smooth as found experimentally, still, the model is only about listening to this musical piece while the EEG experimental data measure the whole brain performing many other tasks at the same time.
	
	Correlations decreases up to 90-100 Hz and increase again to higher frequencies. Oscillations higher than about 100 Hz are not considered within 'classical' bands like delta (0.5-3.5 Hz), theta (4-7Hz), alpha (8-12 Hz), beta (13-30 Hz), or the gamma band which is found at frequencies above 30 Hz\cite{Engel2016}, or split into lower gamma range (30 -60 Hz) and higher gamma range (60-120 Hz)\cite{Freeman2013}. Frequencies above 120 Hz are referred to as fast oscillations\cite{Buzsaki2006} and only briefly associated with conscious content.
	
\begin{figure}
	\centering
	\includegraphics[width=1\linewidth]{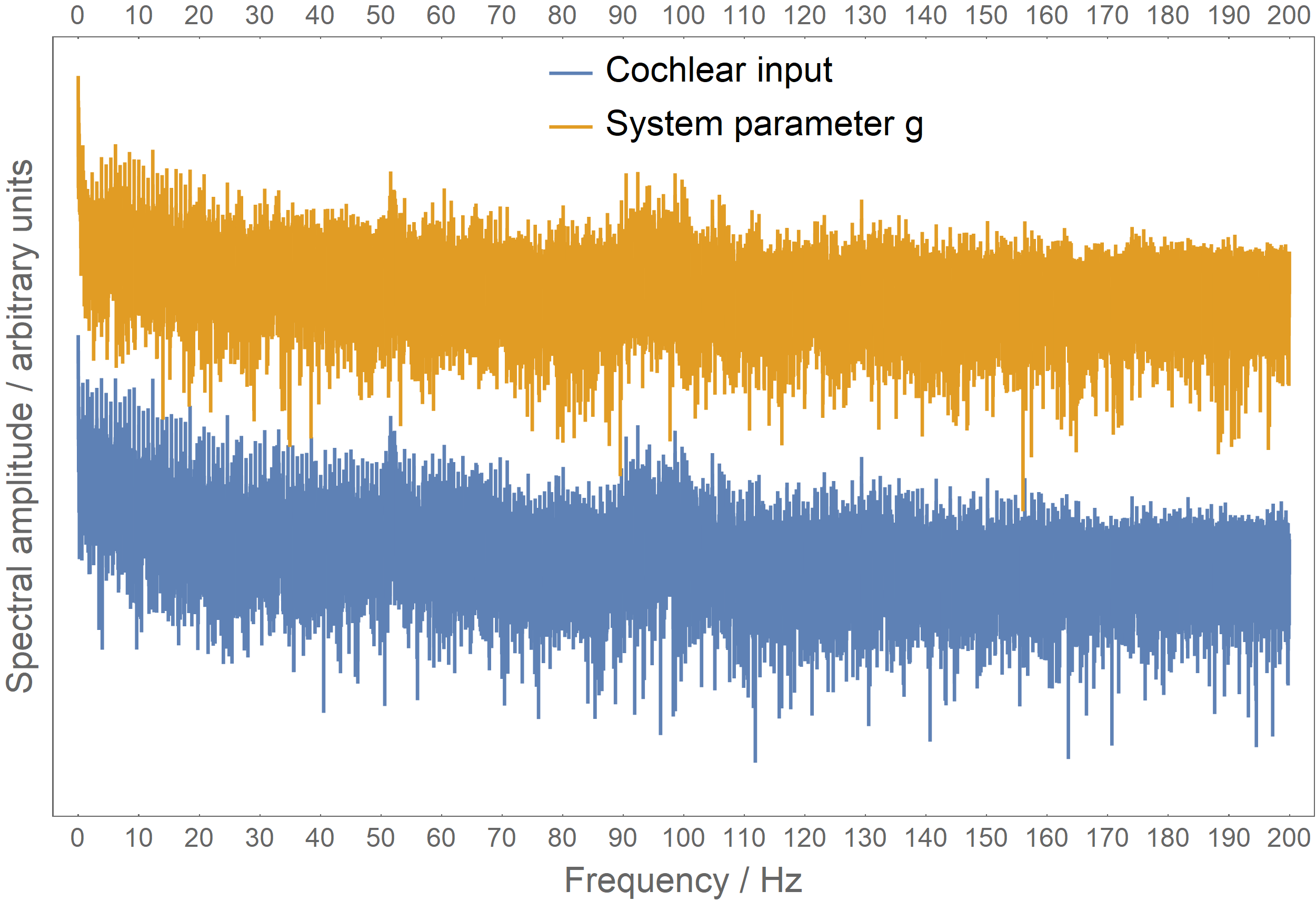}
	\caption{Spectra of the cochlear input $C_t$ and the system parameter g for N=100. Peaks are present at about 50 Hz and 100 Hz. Strong amplitudes at very low frequencies pointing to rhythm content of the musical piece.}
	\label{fig:cochleargspectra100}
\end{figure}

		Looking for an explanation of the gamma band strength, we could expect some connection between the frequency-dependency of $K^C(f)$ and the cochlear input $C_t$ or the system parameter g. Still, connections in terms of their spectral amplitudes are not straightforward. In Fig. \ref{fig:cochleargspectra100} both spectra are shown. Both spectra are very similar, pointing to a response of the Brain IPF to its input, as expected. Indeed, around 50 Hz a peak is present in both time series, corresponding to the 50 Hz peak of $K^C(f)$. Still, the 50 Hz peak in $C_t$ and g are much sharper. Also, $C_t$ and g spectra have a wider peak around 90-100 Hz while the correlation with Kuramoto synchronization has a gap in this region.

		Furthermore, both spectra of $C_t$ and g have enhanced energy in the low-frequency range, pointing to the rhythm content of the musical piece. The piece has 92 BPM corresponding to 1.53 Hz. During the high amplitude sections of the piece, semiquaver notes are played, resulting in 6.1 Hz, during the low amplitude sections the piece uses mainly quaver notes at 3.1 Hz. So much energy is expected below about 6 Hz which is the case.

\begin{figure}
	\centering
	\includegraphics[width=1\linewidth]{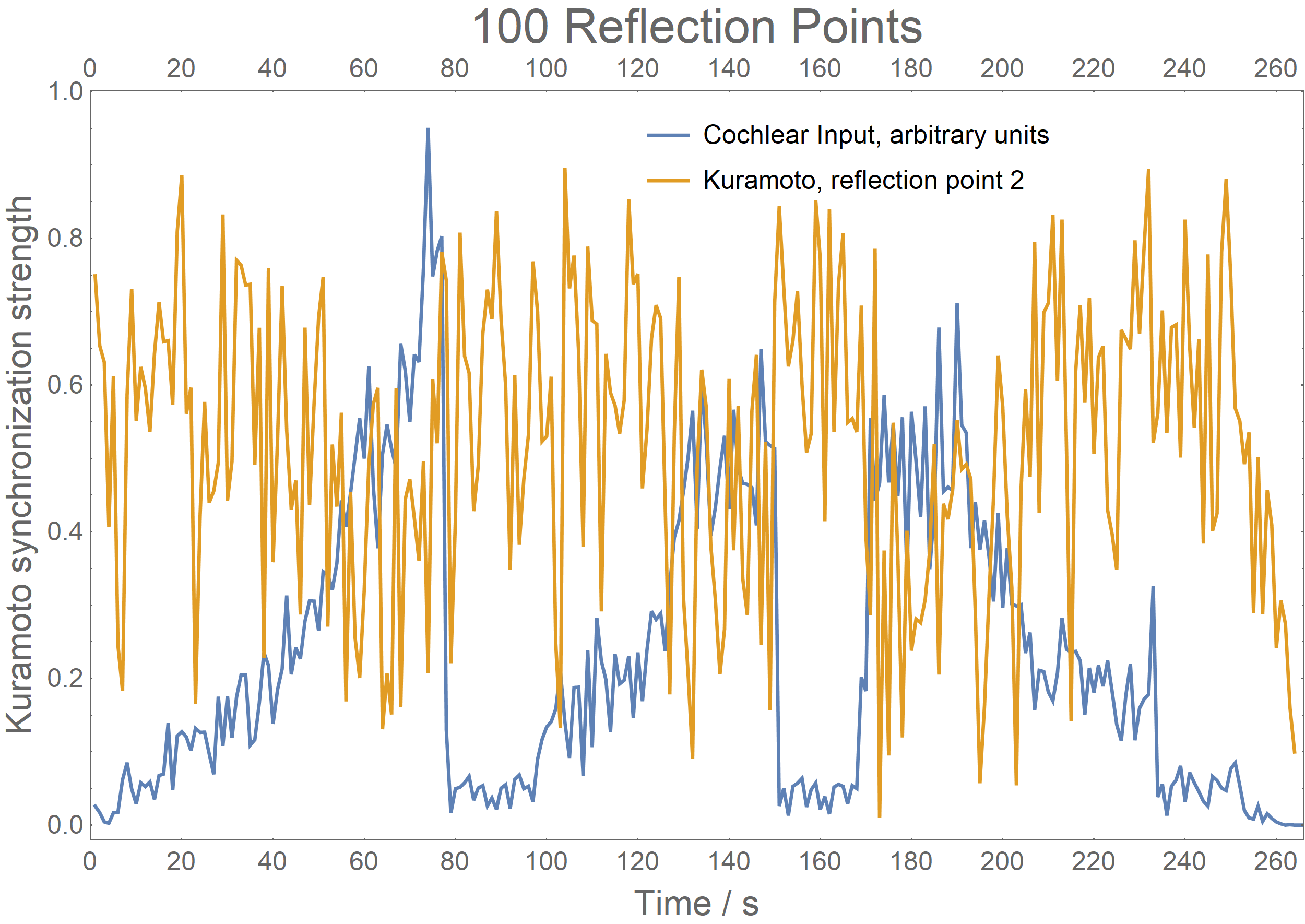}
	\caption{Cochlear input $C_T$ and Kuramoto order parameter $K_T(2 Hz)$ for the N=100 neuron reflection point case. During times of low cochlear input synchronization is very strong, up to 0.8, pointing to a convergence behavior of the system during times of low input at low frequencies, similar to a temporal lobe epilepsy.}
	\label{fig:synchronizationvstime1002}
\end{figure}

	Again, connections to $K^C(f)$ are not straightforward, as about below 10 Hz $K^C(f)$ is mainly negative. This is not found experimentally and is surprising at first. Still when examining an example of N=100 at 2 Hz as shown in Fig. \ref{fig:synchronizationvstime1002} the reason becomes clear. During times of low cochlear input, amplitude synchronization is very strong, while it is lower when the cochlear input is stronger. In a previous study on the principle behavior of the IPF Brain model\cite{Bader2022} it appeared that at a relation of 10\% of inhibitory vs. excitatory neurons in the brain, an optimum in terms of convergence or stability of the system is reached. Therefore, we expect the system to converge when low external input is presented which is the case here. This leads to a systematic increase of synchronization at times of low external input and therefore to a negative correlation.

	For comparison, an IPF Brain model fed with white noise was performed to see if the low-frequency behavior might be a systematic bias of the model. Still, over the whole frequency range low correlations around zero were found with slightly positive and negative values and no frequency-dependency, as expected.

	Such convergence would correspond to epilepsy, as all neurons are synchronized. Indeed, epilepsy is reported in the auditory cortex, associated with creativity or spirituality\cite{McCrae2012}. Such non-everyday experiences are known to occur as trance, often associated with either long periods of isochronous rhythm or during silence. Both is present in the musical piece during times of low cochlear input. Of course, as discussed above, the model is only about listening to a musical piece without performing other tasks like a real brain does. Still, temporal lobe epilepsy is only present at the auditory cortex, and therefore, the negative correlations might model such a case.

	Indeed, in the EEG data, there is enhanced synchronization at such low frequencies again, not found with the FitzHugh-Nagumo model\cite{Sawicki2022}. This synchronization is positive and not negative. This might be due to the EEG data taken from all regions of the human skull, while the IPF Brain model is not specifying brain regions and, therefore, is also suitable to represent the auditory cortex only. Zhe enhanced synchronization at lower frequencies might also be caused by motor action taken in the brain at low frequencies in the body movement dancing range.

	\subsection{Scaling law}
	EEG data of the brain show a scaling law in the brain spectrum. Plotting this spectrum in a log-log plot exhibits a constant slope over a wide range of frequencies\cite{Kozma2016}. This corresponds to a so-called 1/f-noise, as the spectrum follows a $1/f^\alpha$ rule, where a fractal dimension of $\alpha \sim 2$ is found. The scaling law points to a close and systematic connection between the different frequency regions of brain activity and the self-organizing nature of the brain.

\begin{figure}
	\centering
	\includegraphics[width=1\linewidth]{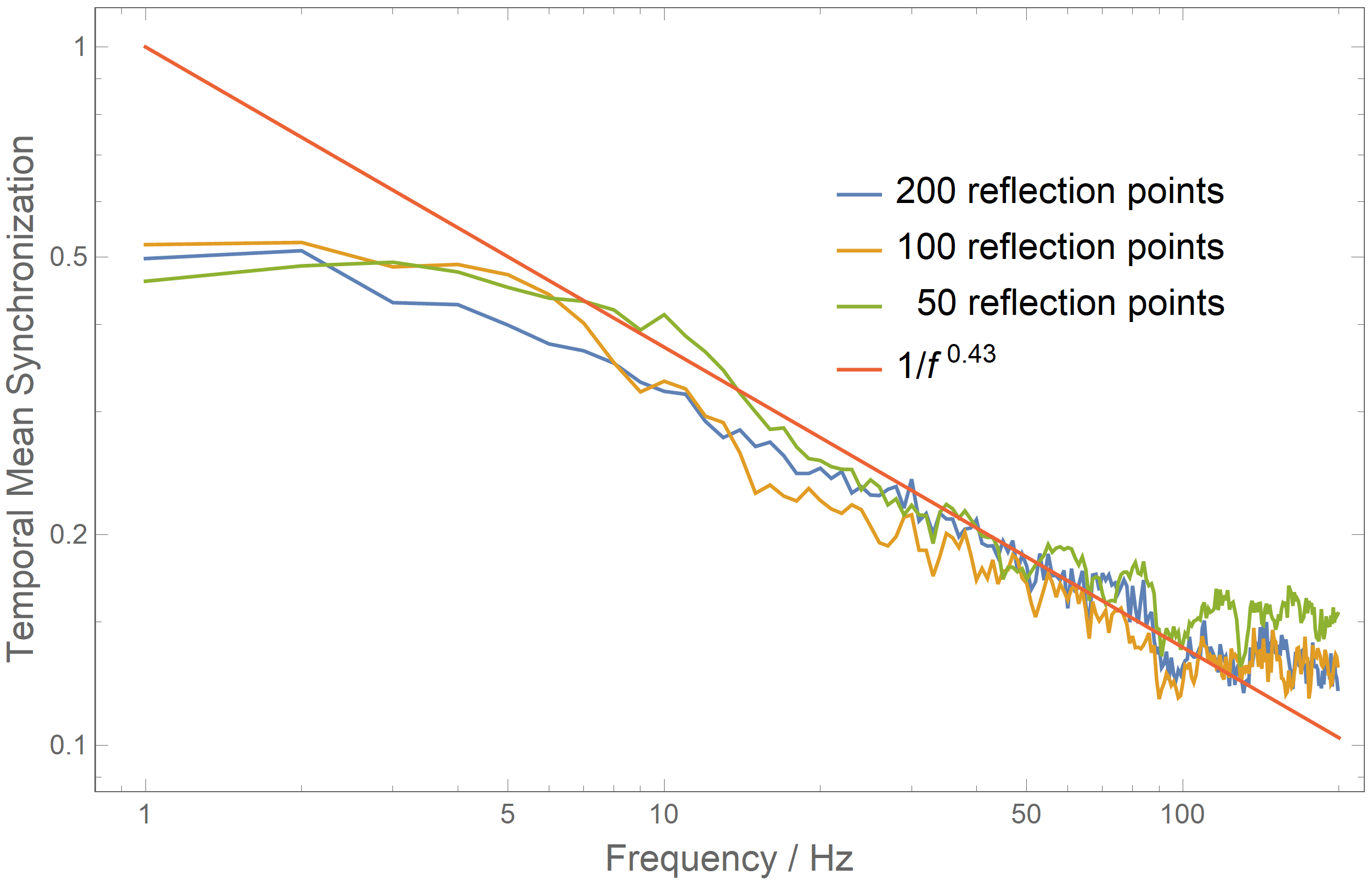}
	\caption{Scaling law of brain synchronization measured as Kuramoto order parameter $K^C(f)$. In a log-log plot over a wide range of frequencies a constant slope is pointing to the scaling law of the IPF Brain model consistent with EEG data.}
	\label{fig:meansynchronization}
\end{figure}

	The IPF Brain model also shows such a scaling law as shown in Fig. \ref{fig:meansynchronization}. Experimental EEG data also show a plateau or even a positive slope up to about 5 Hz, consistent with the IPF Brain findings. The rippling of the spectrum indicates enhanced activity in certain regions like with the gamma band found already above. The only aspect not corresponding to EEG data is the fractal dimension of IPF at $\sim 0.43$ rather than the expected $\sim 2$. The reason here is unclear and subject to future analysis.

		\section{Conclusions and Discussion}
		
		The association of temporal lobe epilepsy with spiritual or meditative experiences corresponds well with the intention of the musical piece \emph{One Mic} used in this investigation. The lyrics report about the hard struggle in criminal gangs with police interaction and shootings. Several references are made to spiritual, especially Christian symbols and ways of similar suffering. Such lyrics are rapped during the large amplitude time frames of the piece, the verses, where also a police siren can be heard. These sections are followed by the low-amplitude parts where the refrain \emph{All I need is one mic} is repeated, pointing to music and lyrics production as an alternative or a weapon against such a hard struggle. These sections contrasts the verses as they are presented in a contemplative or meditative way.

	This compositional tool of presenting a meditative alternative by reducing the volume, reducing the beat from semi-quavers to quavers and omitting most of the sound effects and keyboard pads of the verse is shown in this paper to produce strongly enhanced synchrony and convergence of the neural network. This synchrony is found in the low frequency range where the musical rhythm is represented. 

	Such musical structures are very simple and present in many musical pieces, where regions of low volume and isochronous rhythm are present. Further investigations are needed to model and measure the neural reaction to such musical content in more detail. Still, due to the simplicity of this compositional tool, one can expect composers and musicians to use it in many musical scenarios.

	In a previous similar EEG study of a musical piece \emph{Classical Symphony} by Shemian again, increased brain synchronization is found towards an expectancy point, after which synchronization decreases again\cite{Hartmann2020}. This is a typical EDM piece such that a tension is built up by compositional tools like increased amplitude and event density only to climax at a point where the dense structure ends and a four-to-the-floor bass drum starts. Again, this is a typical compositional tool in EDM, a tension build-up and decay repeated dozens of times during a dance night. There again this compositional tool is clearly present in the EEG data in its large-scale musical form.

	In this study brain synchronization followed the reasoning of a coincidence detection mechanism of cortical oscillators modeling neural activity in the striatum\cite{Buhusi2005}. After a start of neural oscillation increased synchronization peaks at a point a subject expects an event to happen, like waiting at a traffic light to turn green. The oscillation is then expected to include motor regions, making us nervously shaking towards the expected time point.

	This might be considered another compositional tool to make people dance. A tension built-up leading to a neural oscillation and including the motor region enhances the will of subjects to move and, in the case of music, to dance. Although there is strong evidence that this is the case, there is no final proof, as measurements in the motor region are still missing in the musical case.

	This line of reasoning seems fundamentally different than that followed in the present paper of neural synchronization leading to a meditative mood. Still, both cases are confirmed experimentally, also that of the temporal lobe epilepsy\cite{McCrae2012}. The difference might indeed be found in the different frequency ranges, where increased synchronization at higher frequencies, around 50 Hz, might cause a perception of increased tension and synchronization at low frequencies contrary that of a meditative state. Taking into account that synchronization and de-synchronization are fundamental activities in the brain, and brain activity represents all possible states of mind, perception, and consciousness, such differentiation seems plausible.

\section{Acknowledgement}


\begin{thebibliography}{99}
		\bibitem{Baars2013} Baars, B.J., Franklin, S., Ramsoy, T.Z.: Global workspace dynamics: cortical binding and propagation enable conscious contents. Front Psychol 4, 200, 2013. doi:10.3389/fpsyg.2013.00200. 
		
		\bibitem{Babbs2011} Babbs, Ch. F. (2011). Quantitative Reappraisal of the Helmholtz-Guyton Resonance Theory of Frequency Tuning in the Cochlea, \emph{J. of Biophysics,} ID 435135, 1-16.
		
		\bibitem{Bader2022} Bader, R.: Impulse Pattern Formulation (IPF) Brain Model. arXiv:submit/4660893 [q-bio.NC] 21 Dec 2022
		
		\bibitem{Bader2021} Bader, R.: How music works. A Physical Culture Theory. Springer Nature 2021.
		
		\bibitem{Bader2019} Bader, R. (ed.): \emph{Compuational Phonogram Archiving,} Springer Series: Current Research in Systematic Musicology, vol. 5, 2019.
		
		\bibitem{Bader2018a} Bader, R. \& Mores, R.: Cochlear detection of double-slip motion in cello bowing. arXiv:1804.05695v1 [q-bio.NC] 16 Apr 2018.
		
		\bibitem{Bader2018b} Bader, R.: Cochlear spike synchronization and coincidence detection model. Chaos 023105, 1-10, 2018.
		
		\bibitem{Bader2017} Bader, R.: Pitch and timbre discrimination at wave-to-spike transition in the cochlea. arXiv:submit/2066467 [q-bio.NC] 12 Nov 2017.
		
		\bibitem{Bader2015} Bader, R.: Phase synchronization in the cochlea at transition from mechanical waves to electrical spikes. \emph{Chaos} 25, 103124.
		
		\bibitem{Bader2013} R. Bader, \emph{Nonlinearities and Synchronization in Musical Acoustics and Music Psychology,} Springer-Verlag, Berlin, Heidelberg, Current Research in Systematic Musicology, vol. 2,  (2013). 
		
		\bibitem{Blass2020} Blass, M., Fischer, J. \& Plath, N.: Computational Phonogram Archiving: a generic framework for knowledge discovery in music archives, Physics Today 73, 12, 2020.
		
		\bibitem{Blass2019} Bla\ss{}, M. \& Bader, R.: Content Based Music Retrieval and Visualization System for Ethnomusicological Music Archives. In: R. Bader (ed.): \emph{Computational Phonogram Archiving}, Springer Series: Current Research in Systematic Musicology, Vol. 5, 145-174, Springer, Heidelberg, 2019.
		
		\bibitem{Bregmann1990} Bregman, A. S.: \emph{Auditory scene analysis : the perceptual organization of sound.} MIT Press, 1990.
		
		\bibitem{Briot2020} Briot, J.-P., Hadjeres, G. \&  Pachet, F.-D.: Deep Learning Techniques for Music Generation. Springer Nature 2020.
		
		\bibitem{Buhusi2005} Buhusi, C.V. \& Meck, W.H.: What makes us tick? Functional and neural mechanisms of interval timing. Nature Reviews Neuroscience, 6, 755–765, 2005.
		
		\bibitem{Buzsaki2006} Buzs\'aki, G.: \emph{Rhythms of the Brain,} Oxford University Press, 2006.
		
		\bibitem{Cariani2001} P. Cariani, Temporal Codes, Timing Nets, and Music Perception, \emph{J. New Music Research} \textbf{30 (2)}, 107–135  (2001).
		
		\bibitem{Cuddy2018} Cuddy, L. L.: Long-Term Memory for Music. In: R. Bader: \emph{Springer Handbook of Systematic Musicology} 453-459, Springer 2018.
		
		\bibitem{Deutsch2013} Deutsch, D.: The psychology of music. Academic Press series in cognition and perception. Academic, Oxford, 3rd ed. edition, 2013.
		
		\bibitem{Deliege2014} Deli\`ege, I. \& M\'elen, M.: Cue abstraction in the representation musical form. In I. Deli\`ege and J. A. Sloboda, editors, Perception and cognition of music, 387-412. Psychology Press, Hove, 2014.
		
		\bibitem{Hawkins1996} Hawkins, L. H., McMullen, T. H., Popper, A.N. \& Fay, R. (eds.): \emph{Auditory Computation}, Springer Handook of Auditory Research, Springer, New York 1996.
		
		\bibitem{Engel2016} Engel, A. K., \& Fries, P.: Chap. 3-Neuronal oscillations, coherence, and consciousness in The neurology of conciousness (Second Edition)
		(Cambridge, Massachusetts: Academic Press), 49–60, 2016. doi:10.1016/b978-0-12-800948-2.00003-0
		
		\bibitem{Leman1997} Leman, M. \& Carreras, F.: Schema and Gestalt: Testing the hypothesis of Psychoneural Isomorphism by Computer Simulation. In: Marc Leman (ed.): \emph{Music, Gestalt, and Computing. Studies in Cognitive and Systematic Musicology.} Springer, Berlin, 144-168, 1997.
		
		\bibitem{Freeman2014} Freeman, W.: \emph{A socitey of brains,} Psychology Press, 2014.
		
		\bibitem{Freeman2013} Freeman, W. J., \& Quian Quiroga, R.: Imaging brain function with EEG: Advanced temporal and spatial analysis of electroencephalographic signals. New York: Springer, 2013.
		
		\bibitem{Friston2013} Friston, K. J. \& Friston, D. A.: A Free Energy Formulation of Music Performance and Perception. Helmholtz Revisited. In: R. Bader (ed.): \emph{Sound – Perception – Performance,} Springer Series: Current Research in Systematic Musicology, vol. 1, 43-70, Springer, Heidelberg, 2013.
		
		\bibitem{Gerster2020} Gerster, M., Berner, R., Sawicki, J., Zakharova, A., Skoch, A., Hlinka, J., et al.: FitzHugh-Nagumo oscillators on complex networks mimic epileptic seizure-related synchronization phenomena. Chaos 30, 123130, 2020.
		
		\bibitem{Gjerdingen1990} Gjerdingen, R.O.: Categorization of musical patterns by selforganizing neuronlike networks. \emph{Music Perception} 8, 339-370, 1990.
		
		\bibitem{Grossberg1976a} Grossberg, S.: Adaptive pattern classification and universal recording. I: Parallel development and coding of neural feature detectors. \emph{Biological Cybernitics} 23, 121-134, 1976.
		
		\bibitem{Grossberg1976b} Grossberg, S.: Adaptive pattern classification and universal recording II: Feedback, expectation, olfaction, and illusion. \emph{Biological Cybernitics} 23, 187-202, 1976.
		
		\bibitem{Haken2008} Haken, H.: Brain Dynamics, Springer Series in Synergetics, 2. ed., Springer, Heidelberg, 2008.
		
		\bibitem{Hartmann2020} Hartmann, L. \& Bader, R.: Neural Synchroniztaion of Music Large-Scale Form. arXiv:2005.06938v1 [q-bio.NC] 14 May 2020
		
		\bibitem{Hartmann2014} Hartmann, L.: Neuronal synchronization of musical large-scale form: an EEG-study. Proc. Mtgs. Acoust. 22, 035001 (2014); doi: 10.1121/2.0000042.
		
		\bibitem{Kacprzyk2015} Kacprzyk, J. \& Pedrycz, W. (ed.): \emph{Springer Handbook of Computational Intelligence,} Springer, Berlin, Heidelberg, 2015. 
		
		\bibitem{Kozma2016} Kozma, R. \& Freeman, W.J. (eds.): Cognitive Phase Transitions in the Cerebral Cortex - Enhancing the Neuron Doctrine by Modeling Neural Fields, Springer Series Studies, System, Decision, and Control, Springer, Heidelberg, 2016.		
						
		\bibitem{Lerdahl1990} Lerdahl, F. \& Jackendoff, R.: \emph{A generative theory of tonal music.} The MIT Press series on cognitive theory and mental representation. MIT Press, Cambridge, Mass., 4. print edition, 1990.
		
		\bibitem{Linke2022} Linke, S., Bader, R. \& Mores, R.: Describing minimum bow force using Impulse Pattern Formulation (IPF) – an empirical validation. Proceesings of Meetings on Acoustics, in print.
		
		\bibitem{Linke2021a} Linke, S., Bader, R., \& Mores, R.: Modeling synchronization in human musical rhythms using Impulse Pattern Formulation (IPF). arXiv:submit/4062462 [q-bio.NC] 6 Dec 2021
		
		\bibitem{Linke2021b}Linke, S., Bader, R., \& Mores, R.: Influence of the supporting table on initial transients of the fretted zither: An impulse pattern formulation model. Proc. Mtgs. Acoust. 43, 035003 (2021); doi: 10.1121/2.0001494
		
		\bibitem{Linke2019a} Linke, S., Bader, R. \& Mores, R.: The impulse pattern formulation (IPF) as a model of musical instruments—Investigation of stability and limits, Chaos 29, 103109-1, 2019.
		
		\bibitem{Linke2019b} Linke, S., Bader, R. \& Mores, R.: Measurements and impulse pattern formulation (IPF) model of phase transitions in free-reed wind instruments. JASA 146, 2779, 2019, doi.org/10.1121/1.5136626 
		
		\bibitem{Liu2008} S. Liu, R. D. White, Orthotropic material properties of the gerbil basilar membrane, \emph{J. Acoust. Soc. Am.}, \textbf{123 (4)}, 2160–-2171 (2008).
		
		\bibitem{McCrae2012} McCrae, N. \& Elliott, S.: Spiritual experiences in temporal lobe epilepsy: a literature review, British Journal of Neuroscience Nursing, Vol. 8 (2), 2012/13.
		
		\bibitem{Ohl2016} Ohl, F.W.: On the Creation of Meaning in the Brain—Cortical Neurodynamics During Category Learning. In: Kozma, R. \& Freeman, W.J. (eds.): Cognitive Phase Transitions in the Cerebral Cortex - Enhancing the Neuron Doctrine by Modeling Neural Fields, Springer Series Studies, System, Decision, and Control, Springer, Heidelberg, 147-159, 2016.
		
		\bibitem{Ohl2001} Ohl, F. W., Scheich, H. \& Freeman, W. J.: Change in pattern of ongoing cortical activity with auditory category learning, nature 412:733–736, 2001.
		
		\bibitem{Omelchenko2013} Omelchenko, I., Omel\'chenko, O. E., H\"ovel, P., and Sch\"oll, E. (2013). When nonlocal coupling between oscillators becomes stronger: Patched synchrony or multichimera states. Phys. Rev. Lett. 110, 224101, 2013.
		
		\bibitem{Pressnitzer2004} Pressnitzer, D., de Cheveigne, A., McAdams, St. \& Collet, L. (eds.): \emph{Auditory Signal Processing: Physiology, Psychoacoustics, and Models}, Springer, New York 2004.
		
		\bibitem{Sawicki2022} Sawicki, J., Hartmann, L, Bader, R. \& Schöll, E.: Modelling the perception of music in brain network dynamics. Frontiers in Network Physiology 2:910920. doi: 10.3389/fnetp.2022.910920
		
	\end{thebibliography}
\end{document}